\PassOptionsToPackage{bookmarks=false}{hyperref}
\documentclass[10pt, conference, letterpaper]{IEEEtran}

\usepackage[T1]{fontenc} 
\usepackage[utf8]{inputenc}
\usepackage{cite}
\usepackage{amsmath,amssymb,amsfonts}
\usepackage{algorithmic}
\usepackage{graphicx}
\usepackage{textcomp}
\usepackage{balance}
\usepackage{microtype}
\usepackage{subcaption}
\usepackage{units}
\usepackage{enumitem}
\usepackage{url}

\usepackage{breakurl}
\usepackage[hidelinks,breaklinks]{hyperref}
\usepackage{tikz}
\usetikzlibrary{arrows,shadows,positioning,backgrounds}
\usepackage{booktabs}
\usepackage{multirow}
\usepackage{arydshln}
\usepackage{makecell}
\usepackage{dblfloatfix}

\newcommand{\afblock}[1]{\noindent{\textbf{#1 }}} 
\newcommand{\takeaway}[1]{\noindent{\textbf{Takeaway.}} \textit{#1}}
\newcommand{\sref}[1]{Sec.~\ref{#1}}

\usepackage{xcolor}
\makeatletter

\newcommand\thefontsize[1]{{#1 The current font size is: \f@size pt\par}}
\makeatother

\newcommand\copyrighttext{%
  \footnotesize © IFIP, (2018). This is the author's version of the work. It is posted here by permission of IFIP for your personal use. Not for redistribution. The definitive version was published in \emph{Network Traffic Measurement and Analysis Conference (TMA), 2018}, \url{http://tma.ifip.org/2018/wp-content/uploads/sites/3/2018/06/tma2018_paper13.pdf}.}
\newcommand\copyrightnotice{%
\begin{tikzpicture}[remember picture,overlay]
\node[anchor=south,yshift=10pt] at (current page.south) {\fbox{\parbox{\dimexpr\textwidth-\fboxsep-\fboxrule\relax}{\copyrighttext}}};
\end{tikzpicture}%
}

\begin{document}
\bstctlcite{IEEEexample:BSTcontrol}

\title{Demystifying TCP Initial Window Configurations of Content Distribution Networks}

\author{\IEEEauthorblockN{Jan R\"uth and Oliver Hohlfeld}
\IEEEauthorblockA{\textit{RWTH Aachen University} \\
\textit{\{rueth,hohlfeld\}@comsys.rwth-aachen.de}
}
}

\maketitle
\copyrightnotice
\begin{abstract}
Driven by their quest to improve web performance, Content Delivery Networks (CDNs) are known adaptors of performance optimizations.
In this regard, TCP congestion control and particularly its initial congestion window (IW) size is one long-debated topic that can influence CDN performance.
Its size is, however, assumed to be static by IETF recommendations---despite being network- and application-dependent---and only infrequently changed in its history.
To understand if the standardization and research perspective still meets Internet reality, we study the IW configurations of major CDNs.
Our study uses a globally distributed infrastructure of VPNs giving access to residential access links that enable to shed light on network-dependent configurations.
We observe that most CDNs are well aware of the IW's impact and find a high amount of customization that is {\em beyond} current Internet standards.
Further, we find CDNs that utilize different IWs for different customers and content while others resort to fixed values.
We find various initial window configurations, most below 50 segments yet with exceptions of up to 100 segments---the tenfold of current standards.
Our study highlights that Internet reality drifted away from recommended and standardized practices.

\end{abstract}

\section{Introduction}
Content Distribution Networks (CDNs) are a key component of today's Web.
Their ongoing quest to serve web content from nearby servers has flattened the hierarchical structure of the Internet~\cite{labovitz2010internet} and promises lower latencies.
In pursuit of performance, CDNs are known to be early adaptors of new technology in an attempt to optimize the web experience for their customers.
For example, Google has pushed several improvements to web technology, including new protocols such as HTTP/2 (through SPDY) or QUIC, both have found swift adoption~\cite{isthewebhttp2yet,ruethPAM2018QUIC,zimmermannNETWORKING2017push} by other CDNs.
While adopting new technologies offer promising gains, their correct configuration is often challenging---e.g., HTTP/2 server push is regarded as a key feature but known to be notoriously hard to use~\cite{zimmermannNETWORKING2017push,h2PushQoE}.
Some of these configuration challenges stem from the fact that they are depended on network and application characteristics.

One long-debated performance configuration parameter is TCP's (and soon QUIC's) Initial Congestion Window (IW) size.
The IW size controls the amount of unacknowledged data sent at connection start and thereby heavily influences the start-up behavior of new and especially short-lived connections (e.g., typical web transfers) or those that are revived from idle.
A small IW can prolong transmissions and cause unnecessary latency as TCP needs to await feedback (ACKs) to increase the congestion window.
Contrary, too large IWs can lead to loss and retransmissions when the network cannot handle large bursts of data. %
Thus choosing the optimal value for each network is critical for good performance---and interesting for CDNs.

Despite its relevance, the IW size is typically regarded as a {\em static} parameter whose IETF recommended size should fit all networks and applications.
Since its first definition to 1 segment in 1988~\cite{jacobson88}, its recommended size has only changed twice, to 2-4 segments in 1998~\cite{rfc2414, rfc3390} and---motivated by increasing access speeds and shorter page loading times\cite{dukkipati2010}---to 10 segments in 2013~\cite{rfc6928}.
It was very recently shown that IW customization can help in reducing CDN latency~\cite{floresICDCS2016riptide}.
In this regard, a small-scale study by CDNPlanet showed that half of the probed CDNs use IW10 as IETF recommended size while others already use larger IW sizes~\cite{cdnplanet}.
In this regard, others~\cite{allmanDRAFTabandonIW} even argue to abandon static IETF standardized values for the IW to enable customization already in the standards.
Yet, little is known about CDN specific IW configurations.

In this paper, we broadly probe CDNs to gather an empirical understanding on how IW customization already takes place in today's Internet.
By scanning CDN IW configurations, also from globally distributed vantage points, we extend previous work~\cite{ruethIMC17iwv4} and shine a light on the degree of customization that CDNs show today.
Our results show that IW customization beyond standardized practices is already common practice and there exists a gap between standardization, research, and Internet reality.
Specifically, this work contributes the following:
\begin{itemize}[noitemsep,topsep=0pt,leftmargin=9pt]
	\item The first comprehensive analysis of current IW configuration practices of CDNs. We show that IWs are configured up to ten times higher than IETF's current experimental standard.
	\item Further, we observe that CDNs are variable in their IW usage. We find multiple instances of CDNs that deliver data using different IWs for different customers.
	\item	We observe that larger IWs are for example preferred for streamed video instances, yet, content types do not necessarily enforce certain IW settings.
	\item By analyzing IWs through different geographically distributed networks, we find instances of network-dependent IW configurations of CDNs.
	\item We investigate the burstiness of IWs and find that some CDNs utilize pacing to space out packets over time during slow-start to potentially reduce the chance of losses.
\end{itemize}

\begin{figure*}
\centering
\begin{subfigure}[t]{0.5\textwidth}
\centering
\includegraphics{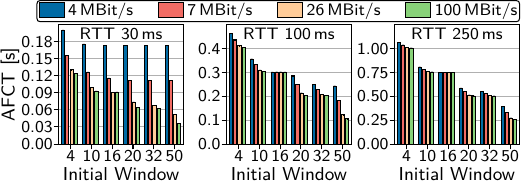}
\caption{Average Flow Completion Time (AFCT) when varying bandwidth, latency, and IWs. Horizontal lines mark roundtrips.}
\label{fig:iw_improvements}
\end{subfigure}~
\begin{subfigure}[t]{0.5\textwidth}
\centering
\includegraphics{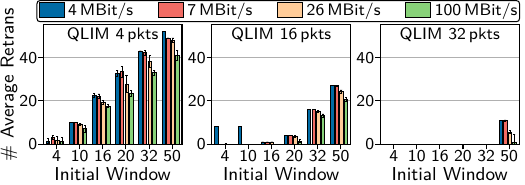}
\caption{Losses increase when increasing IWs depending on bottleneck bandwidth and queue sizes.}
\label{fig:iw_problems}
\end{subfigure}
\caption{Improvements in latency (a) and increases in losses (b) when varying initial congestion windows for different network settings. Benefits of increased initial congestion windows are highly network-dependent.}
\end{figure*}

\afblock{Structure.}
\sref{sec:bg} discusses related work and shows how IWs are defined, standardized, and how they impact performance.
Our IW scanning methodology and architecture is introduced in \sref{sec:scan}.
Following, \sref{sec:results}--\sref{sec:pacing} paints the global CDN IW configuration space %
and \sref{sec:conclusion} concludes our findings.

\section{TCP's Initial Congestion Window}
\label{sec:bg}
We start by exploring TCP's Initial Congestion Window (IW): {\em i)} how it is defined and sized, {\em ii)} how its size influences web performance, and {\em iii)} how related works have gathered an understanding on IWs through Internet measurements.

\afblock{IW Definition.}%
TCP's IW governs the number of unacknowledged bytes in flight until the first acknowledgment is received.
That is typically data sent in the first roundtrip of a connection after the three-way handshake is completed.
Thus, the IW at the sender-side and the receive window at receiver side define the application's data rate at the start of the connection and bootstraps the window doubling during slow start.
Furthermore, depending on the congestion control algorithm, the IW is also used after long idle periods of cached TCP connections for restart (e.g., in web browsers when using HTTP/2).

\afblock{IW Size Definition.}%
The IW is typically defined in bytes and often operating systems allow configuration of the IW in multiples of the Maximum Segment Size (MSS).
To this end, many RFCs (here at the example of~\cite{rfc6928}) define a dualism for the IW, either in terms of the multiple of the MSS, e.g., IW 10 for ten segments worth of data, or by an upper limit of bytes, e.g., 14600 byte typically corresponding to a classic full ethernet frame minus TCP and IP headers times the multiple.

\subsection{Testbed Study: Impact of IW Size on Internet Performance}
To highlight the impact of different IW sizes on Internet performance, we conduct a testbed study.
The testbed involves two directly connected Gigabit Ethernet Linux hosts whose link bandwidth and latency are controlled by NetEm in each host.
We chose four bandwidth configurations (i.e., 4, 7, 26, and \unit[100]{MBit/s}) and three delay configurations (i.e., 30, 100, and \unit[250]{ms}) to reflect typical Internet access configurations reported by Akamai~\cite{akamai_stateoftheinet}.
We further choose six IW configurations of 4, 10, 16, 20, 32, and 50 segments, to reflect the current standard of \unit[4]{segments}~\cite{rfc3390}, the current IETF recommendation of \unit[10]{segments}~\cite{rfc6928}, and larger IWs.
In each experiment, we transfer a single flow of size \unit[71]{kB}, i.e., 50 frames of data (the average size of the Google landing page in 2017). 
For each configuration, each experiment is repeated 30 times.

\afblock{Flow Completion Time.}
Given its relevance to web browsing, we first measure the TCP flows' Average Flow Completion Time (AFCT) subject to the different parameters (i.e., IW size, RTT, and bandwidth).
We define the AFCT as the average time to the last byte of the flow.
The average AFCT for the different parameters and its standard deviation is shown in Figure~\ref{fig:iw_improvements}.
It shows that increasing the IW reduces the AFCT if the link speed or RTT is sufficiently high.
For low bandwidth connections with low latency, larger IWs have effectively no impact on the latency as these connections are limited by throughput.
However, when higher speeds are available, increased IWs can effectively shorten the required roundtrips to finish the data transfer---a key motivation for CDNs to configure larger IWs.

\afblock{Retransmissions.}
Large IWs, however, yield more bursty traffic that can lead to temporary phases of congestion more easily, reflected in higher loss rates.
To highlight this effect, we conducted a second experiment which measures the average retransmission rates of the TCP flows subject to different bottleneck link configurations.
We realize this setting by now connecting the hosts via a bottleneck router with different bandwidth capacities and queue sizes (QLIMs) of a regular drop tail FIFO queue.
We again transfer \unit[71]{kB} and vary the IW configurations for each bandwidth, queue size, and IW triple.
We repeated every configuration 30 times and show the average retransmissions required and standard deviation in Figure~\ref{fig:iw_problems}.

As the figures show, increasing the IW can have detrimental effects on the connection.
We observe that larger IWs cause higher loss rates when either the bottleneck bandwidths or the bottleneck queue sizes are too small.
When we observe the retransmissions for the smallest queue size, we can see that large IWs cause heavy losses.
Increasing the queue size helps in buffering the IWs, yet at the cost of added latency, e.g., a \unit[7]{Mbit/s} link with a buffer of 16 packets can add up to \unit[27]{ms} of delay to a packet.
Thus, simply increasing the queue size is not a desired solution.
While these motivating measurements neglect multiple users sharing the bottleneck, loss-based congestion control of multiple users will lead to full queues all of the time leading to tight buffer space for new flows as shown in these measurements.

\takeaway{
Our study shows, similar to related works~\cite{dukkipati2010,scharfStartupSchemes}, that the IW size can strongly influence flow performance but can also overload congested or low-bandwidth connections.
It is thus key to congestion control to correctly set an initial congestion window that adequately balances throughput and loss to bootstrap a TCP connection.}

\subsection{Related Work}
The relevance of TCP's initial congestion window size is reflected in an extensive debate and a successive evolution of its value in the TCP standards over the last decades.
Initially, the IW was set to 1 segment in 1988~\cite{jacobson88} and 9 years later standardized in 1997~\cite{rfc2001}.
This setting was experimentally extended to 2-4 segments (or \unit[4380]{byte}) in 1998~\cite{rfc2414} and later moved to a proposed standard~\cite{rfc3390}---a setting that remained untouched for a decade.
Motivated by the increase of network access speeds and the desire to reduce web page loading times, \cite{dukkipati2010} proposed in 2010 and later RFC\,6928~\cite{rfc6928} recommended in 2013 to increase the IW to ten segments.
Most recently, Allman~\cite{allmanDRAFTabandonIW} even argues for abandoning a specification of the IW size and thus ending a decades-long debate.
This argument is motivated by allowing hosts to configure more tailored IWs.

Given the relevance of the IW on both flow completion times and Internet traffic burstiness leading to losses, an empirical understanding of the IW is necessary to understand current network performance.
This understanding has been gained in both active and passive measurement studies.
With regards to active measurements, Medina et al.~\cite{medina2005} probed 85\,k servers in 2004 and found most servers to be on an IW of one or two with only 1\% of host having an IW larger than four.
Our measurements are methodological similar to those of Medina, but with a direct focus on CDNs which were still on the rise in 2004 and did not have as much footprint as today.
With regards to passive measurements, Qian et al.~\cite{qian2009} inferred IW distributions from several traces in 2009.
While their dataset covers traces captured in a diverse set of networks and also covers non-publicly visible hosts they did not discuss the impact of CDNs in their study.
A small-scale study by CDNPlanet~\cite{cdnplanet} probed 15 CDNs via HTTP and found 6 to use IW10 and others to use larger IWs.
Our work is similar to that of CDNPlanet in that we share the same goal to shed light on CDN IW configurations, but their methodology did not allow CDNPlanet to grasp a broad assessment of CDN IW configurations which is the focus of this work.
In~\cite{ruethIMC17iwv4} we proposed an approach to estimate TCP's IW for all reachable IPv4 HTTP and TLS hosts which forms the basis of this work.
Our previous approach did enumerate the IPv4 space without host names as a-priori knowledge, thus it cannot measure CDN-hosted sites that use the server name (SNI) within the TLS handshake to indicate which site to deliver.
Many CDNs will only respond with an error paper when no server name is presented, which is often insufficient data to establish an initial window for the probed host and therefore CDNs are not accurately represented in our previous work.
The extension and application to measure CDN IW configuration thus is the focus of this work.

\section{Measuring CDN IW Configurations}
\label{sec:scan}
We next describe our approach to estimate the IW size, its validation, and our overall scanning architecture.

\subsection{Measuring IWs}
\label{sec:measure}
We begin by summarizing our IW size estimation approach.
To enable measuring CDNs, we extend our IW-prober~\cite{ruethIMC17iwv4} to now account for SNI by incorporating per-CDN target URLs and hostnames.
This is needed to fetch large content from CDNs for IW estimation.
We next describe its general procedure and details that we modify from to account for CDN properties.

\begin{figure}
\centering
\scriptsize
\hspace{-.5cm}
\begin{tikzpicture}[font=\sffamily,>=stealth',
commentl/.style={text width=0cm, align=right},
commentr/.style={commentl, align=left}, trim left=(iw)]
\definecolor{bg1}{HTML}{006ba5}
\definecolor{bg2}{HTML}{f26c64}
\definecolor{bg3}{HTML}{ffcc99}
\definecolor{bg4}{HTML}{88d279}

\node[] (init) { Our Prober };
\node[right=2cm of init] (recv) {Probed Host};

\draw[->] ([yshift=-.3cm]init.south) coordinate (syni) -- ([yshift=-.5cm]recv.south) coordinate (synr) node[pos=.5, above,yshift=-0.8mm,sloped] {SYN\,[WIN=65k,WS=4,MSS=...]};

\draw[->] ([yshift=-.3cm]synr) coordinate (synackr) -- ([yshift=-.7cm]syni) coordinate (synackl) node[pos=.5, above, yshift=-0.8mm,sloped] {SYN, ACK};

\draw[->] ([yshift=-.3cm]synackl) coordinate (ackreql) -- ([yshift=-.7cm]synackr) coordinate (ackreqr) node[pos=.5, above, yshift=-0.8mm,sloped] {ACK, GET ... HTTP/1.1};

\draw[->] ([yshift=-.3cm]ackreqr) coordinate (seg1r) -- ([yshift=-.7cm]ackreql) coordinate (seg1l) node[pos=.5, above, yshift=-0.8mm,sloped] {ACK, SEG 1};

\draw[->] ([yshift=-.6cm]seg1r) coordinate (segnr) -- ([yshift=-.6cm]seg1l) coordinate (segnl) node[pos=.5, above, yshift=-0.8mm,sloped] {SEG n};

\draw[dotted] (init.255|-seg1l)--([yshift=-0mm]init.255|-segnl);
\draw[dotted] (recv.285|-seg1r)--([yshift=-0mm]recv.285|-segnr);

\draw[->] ([yshift=-.4cm]segnr) coordinate (seg1retr) -- ([yshift=-.4cm]segnl) coordinate (seg1retl) node[pos=.5, above, yshift=-0.8mm,sloped] {SEG 1};

\draw[->] ([yshift=-.3cm]seg1retl) coordinate (ackl) -- ([yshift=-.7cm]seg1retr) coordinate (ackr) node[pos=.5, above, yshift=-0.8mm,sloped] {ACK n+1, WIN=2 $\cdot $ MSS};

\draw[->] ([yshift=-.3cm]ackr) coordinate (segn1r) -- ([yshift=-.6cm]ackl) coordinate (segn1l) node[pos=.5, above, yshift=-0.8mm,sloped] {SEG n+1};

\draw[->] ([yshift=-.4cm]segn1r) coordinate (segn2r) -- ([yshift=-.4cm]segn1l) coordinate (segn2l) node[pos=.5, above, yshift=-0.8mm,sloped] {SEG n+2};

\draw[->] ([yshift=-.3cm]segn2l) coordinate (rstl) -- ([yshift=-.7cm]segn2r) coordinate (rstr) node[pos=.5, above, yshift=-0.8mm,sloped] {RST};

\draw[->,thick, shorten >=-.5cm] (init) -- (rstl);
\draw[->,thick, shorten >=-.5cm] (recv) -- (recv|-rstl);

\node[commentl, left  =15mm of seg1l.west] {Estimate MSS};
\draw[arrows=|-|] ([xshift=.2cm]seg1r.east) coordinate (toarrtop) -- ([xshift=.2cm]seg1retr.east) coordinate (toarrbot) node [midway, right] {Timeout};
\node[commentr, right  =2mm of seg1retr.east] (to) {Retransmission};
\node[commentl, left  =15mm of seg1retl.west] (iw) {Estimate IW=n};
\node[commentl, left  =15mm of segn2l.west] (iw) {Verify IW full};

\begin{scope}[on background layer]
\path[fill=white!75!bg4] (init.south) rectangle (ackreqr);
\path[fill=white!75!bg1] ([yshift=2mm, xshift=1.15cm]ackreql) rectangle ([yshift=-.1cm]seg1retl -| seg1retr);
\path[fill=white!75!bg1] ([yshift=-.55mm]ackreql) rectangle ([yshift=-.5mm]seg1retl -| seg1retr);
\path[fill=white!75!bg3] ([yshift=-.5mm]seg1retl) rectangle ([yshift=-.5mm]segn2r |- segn2l);
\path[fill=white!75!bg2] ([yshift=-.5mm]segn2l) rectangle ([yshift=-.5mm]rstr);
\end{scope}
\end{tikzpicture}
\caption{Scan procedure: MSS and large receive window are announced and no ACKs are sent until a retransmission. The estimated IW is the \#\,bytes received before the retransmission.}
\label{fig:scanner_procedure}
\vspace{-1em}
\end{figure}
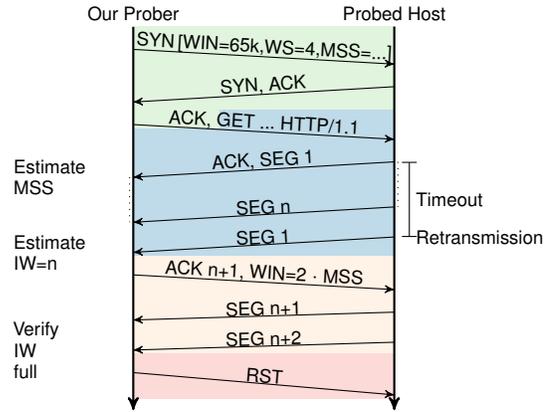

The IW estimation procedure (visualized in Figure~\ref{fig:scanner_procedure}) can be split into four phases:
\textit{First}, a regular TCP handshake is performed announcing a large receive window and, to account for overly large IWs configured at some CDNs, also a window scaling option to never block the data transmission due to flow control.
Since IWs can be configured depending on the MSS, we additionally set a MSS (varied by the later measurements).
No further options like Selective Acknowledgements, e.g., causing TCP tail-loss probes that would challenge IW estimation are activated.
After establishing the TCP connection, the \textit{second} phase starts by transmitting an HTTP GET request in hope to trigger a response that exceeds the configured IW at the probed host.
The probed host will now commence sending the requested resource, however, we are not going to generate acknowledgments for any segment that we receive, thus the initial congestion window will not increase and the host can only send as many bytes as the IW.
By not acknowledging segments, the probed host will eventually initiate a retransmission of the first (from its point of view) lost segment, which heralds the start of the \textit{third} phase.
Either, the sending host was in fact limited by the IW or it ran out of data to send.
To test for this, we start acknowledging the last segment enabling the host to continue sending data and if the host does so we know that the host did not run out of data.
At this point, we are able to estimate the IW by observing the sequence number space and segment sizes that we received before the retransmission.
Finally, the \textit{last} phase consists of tearing down the connection with TCP's RST mechanism.
As this methodology is prone to tail-loss, it is recommended to perform multiple scans of the same host.

\afblock{Implementation and Validation.}%
We implement our tool~\cite{iwCDNSource} in go-lang to benefit from its multiprocessing capabilities.
To test our implementation and to validate the correctness of the IW estimation, we test our tool in mininet~\cite{mininet}.
We use iptable's statistic module to drop packets at the head, within, and at the tail of the IW to validate the estimation correctness, i.e., correct estimations for the first two, and a reduced IW for the latter case (tail-loss).
Further, we vary the IW configuration and available bytes on the server-side and the announced MSS at the probing client to validate non-standard IWs and out-of-data situations in various settings.
Out tool always provided correct IW estimations except for tail-loss (as expected).

\subsection{Measuring CDN IWs}
\label{sec:iwmethodology}
Our measurement study is structured into three phases. At first, we gather lists of target URLs that are served by CDNs. In a second phase we derive the initial windows of the hosts serving the URLs. At last, we use VPNs to derive the IWs from different networks for a subset of these URLs.

\afblock{Target Addresses.}
The first phase is relatively straightforward, we utilize data published by the HTTP Archive~\cite{httparchive}.
The HTTP Archive crawls websites while recording diverse information about the websites, for their bi-monthly crawls they visit all websites included in the Alexa Top 1M list.
We utilize the crawl data from the 15\textsuperscript{th} of January 2018 and extract all URLs that are loaded throughout the crawl, i.e., the landing page URLs as well as objects that are subsequently requested such as images or Javascripts.
Even though the HTTP Archive already marks CDNs in their data, we repeat this step as the CDN choice could be geo-location dependent and as the HTTP Archive data can be up to half a month old the CDN operator could have changed in the meantime.
To do so, we apply the domain list~\cite{wptcdndns} published by the WebPagetest~\cite{webpagetest} framework (the framework driving the HTTP Archive), this list enables to classify a URL by resolving its domain using DNS.
Many CDNs utilize the DNS to redirect (using CNAME records) a user to the CDN server.
Thus a CDN can be identified by its CNAME pattern in the DNS resolution step.
The result is a rather large list of URLs which we filter to include only URLs hosted at CDNs and only one URL per domain. For each domain, we choose the URL with the largest object size.
This results in a list of $\approx$ 227k URLs hosted on 69 CDNs for which we are going to establish initial windows.
116K objects (25 CDNs) that are too small to reliably estimate an IW (for large segment sizes, see \sref{sec:results}) are removed from the results.

\begin{figure}
\center
\includegraphics{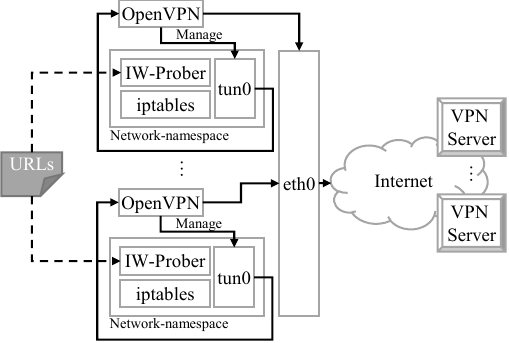}
\caption{Overview of our scanning architecture. We leverage Linux network namespaces to scan concurrently and to easily manage multiple VPN connections.}
\label{fig:arch_overview}
\vspace{-1em}
\end{figure}

\afblock{Scanning Architecture.}
We use the architecture depicted in Figure~\ref{fig:arch_overview} to structure and perform our scans.
To enable concurrent scanning at multiple vantage points, we make use of OpenVPN and Linux's network namespaces.
A network namespace can be seen as a shallow copy of the network stack with its own interfaces and routing tables.
As many VPNs apply Network Address Translation (NAT) to assign IP addresses to their peers, we experienced that different VPNs assigned the same IP or the same subnet to us.
To overcome this issue, we override OpenVPN's device creation and insert a script to manually create network devices in a new network namespace identified by the VPNs publicly facing IP.
This enables to completely disregard any routing or name clash issues when using multiple VPNs in parallel.
We then start one instance of an IW-prober in each namespace and feed it with the URLs.

To not put a large burden on the VPNs, we perform a preprocessing step.
Instead of querying all 111k URLs (potentially multiple times to account for tail losses) through the VPNs we first derive a list of candidate URLs locally.
We select query candidates by grouping URLs hosted at the same CDN using the same IW (derived locally) and select a random sample of URLs for each (CDN, IW) pair.

\afblock{Vantage Points.}
Gathering globally distributed vantage points that grant packet-level access is hard.
To do so for our measurements, we make use of the VPN Gate~\cite{vpngate} project by the University of Tsukuba.
This project's goal is to give access to the Internet without censorship.
To this end, the project manages a list of thousands of relay VPN servers around the globe, many of them are operated by volunteers via their private Internet uplinks.
While the site lists many VPNs, we found only a small set of them to reliably work for our measurements which might also be due to the implemented censorship protection announcing false gateway servers.
To account for our scan methodology, we use only VPN connections made through TCP, thus loss between our VPN client and the VPN server is automatically resolved and is not accounted as loss for our prober.
According to~\cite{vpngate}, most of the VPN servers only have a relatively small bandwidth capacity mostly below \unit[10]{MBit/s}.
Consequently, to not disturb the regular VPN operation, we implement a rate shaper into our prober that smoothes burst and limits the outgoing bandwidth.
We configure it to transmit at most 100 packets per second, thus we limit the prober to \unit[$\approx$1.2]{Mbit/s} for full-sized frames and much less for smaller frames.
Further, through local experiments we found that excessively parallelizing IW estimations challenge NATs easily causing exhausted NAT tables, therefore, we limit ourselves to a handful of parallel estimations per VPN.

\section{Campus Network Perspective on CDN IWs}
\label{sec:results}

We next explore CDN IW configurations from the perspective of a well-provisioned campus network (RWTH Aachen University) (worldwide perspective follows in \sref{sec:vpniw})
to set an upper bound on the expected IW sizes.
As our network's upstream ISP peers with DE-CIX (at which many CDNs peer as well) and our networks offers at least one order of magnitude higher capacity than typical consumer Internet connections, CDNs could potentially adapt by serving content with higher IWs thus providing an upper bound on the expected IW sizes.

\afblock{IW Probe Procedure.}
As IWs can be configured in bytes or segments, we scan each URL (see \sref{sec:iwmethodology}) with different maximum segment sizes of 64, 128, 536, 1200 bytes, ten times each.
This enables to derive if the scanned host changes the total number of bytes delivered in the IW, i.e., the IW is fixed to a certain number of packets (we refer to the segments) or if it is fixed to a certain amount of bytes (we refer to the bytes).
To account for tail-loss, we perform a majority vote for each segment size and regard a scan as successful if \unit[$>$]{\unit[50]{\%}} of the votes agree on the largest observed IW (97\% of measurements).
To derive the final IW, we inspect the number of packets and bytes received over the four different segment sizes: if the IW depends on the segment size, we calculate an IW (in bytes) as if we were using maximum-sized segments (\unit[1460]{byte}), otherwise, we directly use the fixed amount of bytes.
Note that we refrain from showing quantities in which we observed certain IWs as they could be biased by the choice of URLs.
Furthermore, we are not able to estimate IWs for all URLs, since their object size can simply be too small fill a larger IW. This would bias the results towards smaller IWs.

\subsection{IW Sizes}
\begin{figure}
\center
\includegraphics[]{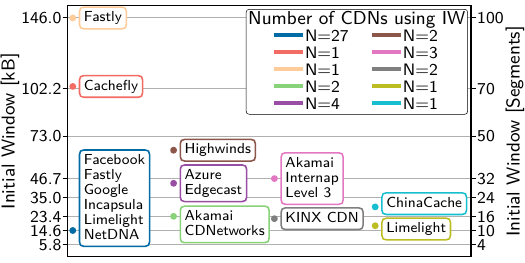}
\caption{CDN IWs as seen from our university network. IETF sizes 4 (not shown) and 10 are present but also larger IWs.}
\label{fig:ac_iws}
\vspace{-1em}
\end{figure}

Figure~\ref{fig:ac_iws} shows the resulting IW sizes in bytes and segments assuming \unit[1460]{byte} packets from our local campus network.
Each dot represents an IW configuration, the adjacent box lists a selection of CDN providers that deliver URLs with this IW (a CDN can occur in multiple boxes).
Even though we find many CDNs offering URLs via IW10, we also find much larger sizes.
This is in contrast to our prior IP only scans over the IPv4 address space~\cite{ruethIMC17iwv4}, which found IETF recommended IW sizes to dominate  given the number of legacy systems (e.g., DSL gateways).

Our findings show that CDNs do in fact depart from IETF recommended IW sizes and customize the IW.
For example, we observe IW16 and IW32 for the probed Akamai URLs\footnote{We remark that each CDN can use {\em additional} IW configurations beyond the configurations discovered in our measurements.}, both larger than the current IETF recommendation of IW10.
However, we also find very large IWs.
For example, the largest IW that we observed is by Fastly, they deliver {\em some} URLs using an IW of 100 segments.
This large IW has already caused drops during our measurement which might be an indicator that IW100 is too large.
Cachefly also shows a larger than usual IW of \unit[105]{kB}, notably, Cachefly uses a fixed IW configured in bytes which leads to many transmitted segments when small segment sizes are used.
On the opposite end of the spectrum, we find URLs hosted on CDNs that deliver data with a smaller IW than currently recommended.
For example, we find URLs hosted on ChinaCache (not shown) that are delivered with an IW of 4, yet, we can again observe that ChinaCache customizes as well, as they also deliver URLs with IW20.

\begin{table}[t]
\centering
\begin{tabular}{@{}lllll@{}}
\toprule
Operating System & RWIN [B]  & WS & WIN [B] & Segs.\\
\midrule
Linux 4.4 & 58 & 512 & 29.696 & 20\\
Android 6.0 (Linux 3.4) & 685 & 128 & 87.680 & 60 \\
Android 7.0 (Linux 3.18) & 641 & 128 & 82.048 & 56\\
iOS 11.2.5 & 2.058 & 64 & 131.712 & 90\\
Mac OS 10.9.5 & 8.235 & 16 & 131.760 & 90\\
Mac OS 10.13.2 & 4.117 & 32 &  131.744 & 90 \\
Windows 7 (SP 1) & 256 & 256 & 65.536 & 44\\
Windows 8.1 / 10& 1.024 & 256 & 262.144 & 179\\
\bottomrule
\end{tabular}
\caption{TCP receive window (RWIN), window scaling (WS), resulting window (WIN) in bytes and full-sized segments on different operating systems as reported on an HTTP GET request from an otherwise idling system.}
\label{tab:initrwin}
\vspace{-1em}
\end{table}

\afblock{Can Increased IWs be Utilized?}
The actual amount of data that is transported is of course not only dependent on the server's congestion window.
The client permanently announces a receive window (RWIN), TCP demands that no more than the minimum of the advertised RWIN and the congestion window is in flight.
Table~\ref{tab:initrwin} shows the client's advertised receive window on an HTTP GET request for a selection of client operating systems.
As the table highlights, the largest IWs that we measured would not be effective for a couple of operating systems.
Linux 4.4 shows the lowest advertised receive window which would not be able to utilize many of our discovered CDN IWs.
We found a git commit~\cite{gitInitRWND} documenting this receive window in response to the IW10 increase.
Interestingly, Android, even though using an older Linux kernel, has increased the receive window and would be able to utilize most of the IWs measured, the same holds for iOS and all other tested Mac OS variants.
Apart from Windows 7, all recent Windows variants announce receive windows large enough to not thwart even the largest observed IW.

\takeaway{
We observe CDNs to configure IW sizes beyond IETF recommended values, highlighting that i) Internet reality departed from standardization and ii) and IW sizes larger than standardized are practically relevant. 
Their actual impact on network performance, in terms of losses, fairness and flow completion is practically unexplored by current research, highlighting that Internet reality also departed from research.
We found that CDNs do customize IWs, however, it remains unclear when a CDN decides to utilize which IW.
}

\subsection{Are IWs Content Dependent?}
One way to customize IW sizes is by delivery service class (e.g., low latency web delivery vs.\ elastic download), which can explain multiple observed IWs per CDN.
Since we cannot directly identify service classes, we analyze IWs for typical {\em content types} by filtering the HTTP Archive for Akamai-served URLs according to their mime type.
We focus on Akamai, as one of the largest CDNs for which we already observed multiple IW sizes.
For each domain, we take the largest URL of the following mime types: {\em i)} \texttt{application/mp2t} (\unit[62]{URLs}) typically employed for streamed video streaming applications, {\em ii)} \texttt{image/png} (\unit[1812]{URLs}) for images, {\em iii)} \texttt{application/javascript} (\unit[1395]{URLs}) for regular website content, and {\em iv)} \texttt{application/octet-stream} (\unit[67]{URLs}) for any binary data (download).
We expect that interactive content uses the larger of the two initial windows as e.g., the play-out of a video should start as fast as possible.

\begin{figure}[t]
\center
\includegraphics{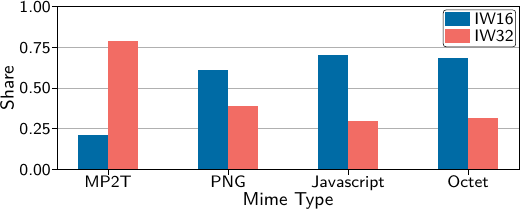}
\caption{IW distribution for Akamai URLs per mime type.}
\label{fig:iw_ak_mime}
\end{figure}

Figure~\ref{fig:iw_ak_mime} visualizes the analysis.
Our expectations are partially met, i.e., streamed video content (MP2T) is in fact mostly delivered with IW32, yet not exclusively.
This and also the other mime types highlight, the mime type does not determine the initial window per se.
For PNGs, Javascripts, and binary data we observe that the majority is served via IW16, the quantity of IW32 varies between \unit[30]{\%} (Javascript, binary) and \unit[40]{\%} (PNG).
These observations highlight that it is more likely that an IW is not set depending on the mime type but is rather dependent on the service class (product) that has been purchased at the CDN.
Of course, some products are designed for interactive delivery and others not, yet, in the end, this non-strict assignment of IWs to mime type shows that the customers decide what they deliver through which product.

\takeaway{
Different content types can benefit from different IW sizes and our results suggest that content dependent customizations (e.g., for interactive video streaming) exist.
Yet, they cannot be purely detected by mine type since they rather depend on the delivery strategy selected at the CDN.
}

\section{Worldwide Perspective on CDN IWs}
\label{sec:vpniw}
To investigate if CDNs tailor IWs to networks, we probe the same URL from multiple vantage points.
To do so, we utilize the public VPNs listed at VPNGate~\cite{vpngate}.
As the service lists thousands of VPNs, we concentrate on a small subset of 14 VPNs all located in different countries and ASs.
For these VPNs, we test samples of URLs (5 per IW/CDN combination) for which we have already established an IW locally, thus enabling to compare if other networks are subject to different IW configurations.

\begin{table}[t]
\centering
\begin{tabular}{@{}lllll@{}}
\toprule
\#VPN & AS & AS Name & Country & Link Type \\
\midrule
1 & AS1221 & Telstra & Australia & Consumer\\
2 & AS3303 & Swisscom & Switzerland & Consumer\\ %
3 & AS3326 & Datagroup & Ukraine & ? \\ 
4 & AS4766 & Korea Telecom & South Korea & ? \\ %
5 & AS7552 & Viettel & Vietnam & Consumer\\ %
6 & AS7922 & Comcast & USA & Consumer\\ %
7 & AS9198 & Kaztelecom & Kazakhstan & Consumer\\ %
8 & AS12389 & Rostelecom & Russia & Consumer\\ %
9 & AS16276 & OVH & France & Datacenter\\ %
10 & AS17534 & NSK & Japan & ?\\ %
11 & AS24560 & Airtel & India & Consumer \\ %
12 & AS24620 & Riga Tech. Univ. & Latvia & University\\ %
13 & AS28548 & Cablevisi{\'o}n & Mexico & Consumer\\ %
14 & AS28885 & OmanTel & Oman & Consumer\\ %
\bottomrule
\end{tabular}
\caption{Classification of VPNs used to estimate CDN IW configurations.}
\label{tab:networks}
\end{table}

Table~\ref{tab:networks} gives an overview of the VPN locations (as reported by VPN Gate), networks as well as a manual classification of their link's nature.
We classified the link type by inspecting {\em i)} the AS and {\em ii)} the reverse DNS name of the VPN host and check if it includes keywords such as: cable, (A)DSL, dynamic, etc.
Most of our VPNs are located in residential access networks, with the exception of one datacenter (\#9), one university network (\#12) and three links (\#3, \#4, \#10) that could not be classified due to missing hints.

\begin{table}[t]
\centering
\setlength\tabcolsep{1.4pt}%
\begin{tabular}{@{}c|cc|c|c|c|c|cc|c|c@{}}
\toprule
\multirow{2}{*}{VPNs} & \multicolumn{2}{c|}{Akamai} & Azure\ & Cachefly& Cloudf. & Edgecast  &\multicolumn{2}{c|}{Fastly} & Highw. &Level\,3\\
           & 16 & 32 & 30 & \unit[105]{kB} & 25 & 30 & 100 & 10 & \unit[64]{kB} & 32\\
\midrule
1,9 & 1 & 1 & 1 & 1 & 1 & 1 & 1 & 1 & 1 & 1 \\
2-6,8 & 16 & 32 & 30 & 105kB & 25 & 30 & 61-62 & 10 & 64kB & 32\\
7 & 16 & 32 & 20-30 & 105kB & 25 & 20-30 & 61-63 & 10 & 20-60kB & 32\\
10 & 16 & 32 & 30 & 105kB & 25 & 30 & 1-100 & 10 & 83kB & 32\\
11 & 16 & 32 & 30 & 105kB* & 25 & 15-30 & 2-61 & 10 & 4-49kB & 32\\
12 & 16 & 32 & 30 & 105kB & 25 & 30 & 87-99 & 10 & 64kB & 32\\
13 & 16 & 32 & 6-30 & 105kB & 25 & 2-30 & 6-73 & 10 & 64kB & 32\\
14 & 16 & 32 & 30 & 105kB & 25 & 30 & 61-75 & 10 & 58kB & 32\\
\bottomrule
\end{tabular}
\caption{IW configurations observed at the VPNs. The top row shows the CDNs with their IWs as discovered within our campus network. Each field marks the IW we discovered through the VPN or a range if we saw consistent losses. Results marked with (*) experienced packet loss but no tail-loss.}
\label{tab:iw_vpn_results}
\end{table}

\begin{figure*}[t]
\center
\includegraphics[]{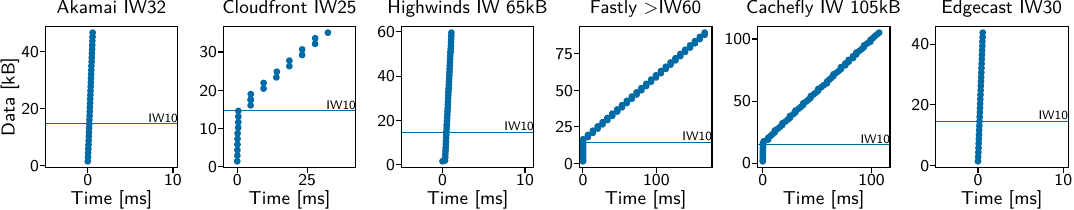}
\caption{IW burstiness for a subset of the observed CDNs and URLs (each with an RTT of \unit[60]{ms}-\unit[70]{ms}). The arrival time of full-sized \unit[1500]{byte} packets (dots) in the entire IW is shown on the x-axis, the IW size (in kB) on the y-axis (e.g., IW10 = \unit[15]{kB}). Note different axis scalings due to different IW sizes.
Some CDNs seem to utilize packet pacing while others do not.}
\label{fig:ac_pacing}
\vspace{-1em}
\end{figure*}

Table~\ref{tab:iw_vpn_results} summarizes our IW estimations through these VPNs.
We were able to build classes of VPNs that perform similarly, already indicating that many of our VPNs show a similar performance and we see a similar IW configuration.
The first class for VPNs \#1 and \#9 show the largest divergence from our campus network.
Here we measured an IW of only 1 segment for all CDNs contacted via both for the consumer (\#1) and for the datacenter (\#9) link.
Especially for a datacenter link this seems too low and does not fit the rest of our data.
When more closely inspecting both VPNs, we found that both VPNs seem to heavily rate-limit the packet-rate.
Even when performing a regular download of the URLs, we are unable to ever get more than two segments in a roundtrip at any time.
Thus, we believe that the IW estimation here is unable to determine the actual IW due to the rate-limiting which highlights the challenges when using vantage points that are out of direct control.

The second, largest class, of VPNs, paints a similar picture to that of our local observations.
We observe for all but one CDN provider the same IWs as seen from our university network.
The only difference being Fastly, for which we have measured IWs between 61-62, we consistently measured IWs in that range indicating that our measurements are subject to loss.
We take this as an indication that it is likely that 62 is not the actual IW that should have been delivered but rather a larger IW was subject to heavy tail-loss, especially since all other IWs are configured similarly to our local observations.

This impression continues when observing the remaining VPNs, there the IWs for Fastly also reach up to 100 segments (VPNs \#10 and \#12) but with consistent losses between multiple measurements.
For many, we observe IWs in the range of 60 to 70 segments.
We take this as an indication of a service specific configuration rather than a network-dependent one.

But we also find patterns of network-dependent configuration, e.g., for the Highwinds CDN that we measured with an IW of 64kB locally.
For VPNs \#10 and \#14, we observe different IWs consistently.
For VPN \#10 we observe a larger IW of \unit[83]{kB} and for \#14 only \unit[58]{kB}.
Yet, also for Highwinds, we can observe losses at VPN \#7 and \#11.

Especially, VPN \#11 observes the highest losses throughout our measurements.
Here, also Cachefly with the second largest IW (equalling to 72 full-sized segments) that we observed shows losses (which does not show any losses at other VPN).

\takeaway{
Overall, we observe that many CDNs use the same IWs regardless of the network and are successful in delivering it without losses.
Interestingly, we find that Level 3 and Akamai both deliver content with IW32 and are successful in delivering it while e.g., Edgecast and Azure with IW30 show losses over the same links.
}

Motivated by these observed losses, we want to investigate the burstiness of IWs.
To this end, a recent proposal~\cite{allmanDRAFTabandonIW} recommends using TCP pacing to evenly space out packet delivery over the RTT when exceeding an IW of 10 segments to be less aggressive towards queues.
This has also been proposed in~\cite{pacingafteridle} after idle slow-start restarts.
Since Linux Kernel 3.11 (released in September 2013), it offers pacing support via a special packet scheduler in the traffic control subsystem, starting with Linux Kernel 4.13 (released in September 2017) also directly from within the TCP stack.
Thus, we continue to investigate the temporal characteristics of the packets transmitted in the IW to investigate the use of pacing at CDNs.

\section{Burstiness of the CDN IWs}
\label{sec:pacing}
To investigate the use of TCP pacing by CDNs we again focus on our university network as we require fine-grained packet arrival times which are not preserved through the VPNs.

\afblock{Pacing Realization.}
The Linux pacer's default configuration works as follows (at the start of a connection): during the three-way handshake, the RTT is estimated as the difference between the SYN and the corresponding ACK. TCP then calculates a pacing rate as the ratio of the current congestion window (in this case the IW) and the estimated RTT which results in a rate at which data will be scheduled to leave the system.
The pacer enforces the limit by delaying packets, however, it allows a configurable initial burst of ten full frames and the subsequent frames are sent in trains of two packets (also configurable).
Furthermore, the pacer can be configured to be more or less aggressive during slow-start or congestion-avoidance by passing a sysctl parameter that is multiplied with the estimated pacing rate.
Thus the expected outcome at the connection start is a burst of ten packets and following trains of two packets.
We next empirically probe CDNs for this pattern to detect pacing.

\afblock{Measuring the Packet-Pacing.}
To measure if the CDNs utilize pacing, we take a look at the packet arrival-times when executing an IW scan.
To do so, we simply record packet traces (with tcpdump) but instruct our IW-prober to delay the ACK following the SYN/ACK by roughly \unit[50]{ms} to emulate a larger RTT to the measured CDN.
We found that the packet coalescing of the NIC that reduces the interrupt-rate causes imprecise software timestamps of the arriving packets, we thus instruct tcpdump to configure our NIC to perform hardware timestamping at packet arrival.

Figure~\ref{fig:ac_pacing} shows all packets (dots) sent during one initial window after connection start for a selected subset of CDNs and URLs.
Their arrival time is depicted on the x-axis and the IW size in kB on the y-axis.
Please note the different x- and y-axis scaling due to the different IW sizes.
We can visually observe two different patterns.
The first, here presented by Akamai, Highwinds, and Edgecast, shows close to no temporal distribution of packets.
The second, presented by Cloudfront, Fastly, and Cachefly, shows a stream of packets arriving virtually at the same time followed by a temporally skewed train of other packets.
The latter follows the expected output of Linux's packet pacer as described before.
This is best visible in the example of Cloudfront, were a burst of ten segments is almost perfectly followed by delayed trains of two packets.
Thus, we can see that some CDNs are likely utilizing pacing during slow-start for IWs larger than IW10 as recommended in~\cite{allmanDRAFTabandonIW}.

When looking at the two largest IWs that we observed by Cachefly and Fastly\footnote{Please note that while measuring pacing, we experienced heavy tail-loss with Fastly leading to the reduced bytes.}, we can see that both pace their IWs, however, we observe that Cachefly is more aggressive in doing so as they spread their IW over roughly 1.5x the RTT while Fastly does it over roughly 2.5x the RTT.

\takeaway{
We find it is likely that pacing is used by some CDNs.
In fact, the two largest IWs show clear pacing patterns.
Past research suggests that pacing can help to bootstrap new or idle connections, however, there is currently only a limited understanding on the impact of pacing on networks as well on its benefits and drawbacks especially as current pacers deviate from a perfectly paced packet streams found in literature.
}

\section{Conclusion}
\label{sec:conclusion}
This paper's goal is to better understand the current configuration of TCP's initial congestion window (IW) by CDNs.
The IW is a long-debated performance parameter.
Its size is in principle network and application dependent, where too small IWs can add unnecessary latency and too large IWs can cause congestion and thus loss.
Yet, the IW is regarded as a {\em static} parameter that fits all networks and applications.
Its IETF recommended size changed only infrequently in its history.

We find that CDNs are well ahead of current IETF standardized practices by using {\em custom} IW configurations.
In our measurement study, we observe IW configurations that are up to ten times higher than the most recent experimental standard.
Our results suggest that CDNs do customize IWs for different services or customers, yet while advantageous for some content types, the content type does not enforce the IW.
On a larger scale, we survey if CDNs adjust IWs depending on the end-user's network.
We find some CDNs for which we can show that IWs vary depending on the network, but not for all.
Driven by losses in our measurements, we analyze the burstiness of the IW delivery and find that some CDNs utilize pacing to space out packets over time.
The largest IWs in our study utilize this feature, yet there is no clear indication if pacing enables these large IWs as the benefits or drawbacks of pacing require further research.
While our study focusses on TCP, QUIC borrows TCP's congestion control and startup phase including initial windows highlighting its future relevance (also in light TCP-BPF~\cite{tcpBPF}).
We thereby aim to inform standardization and academia about current CDN practices that depart from current knowledge and IETF recommendations.
We posit that further research needs to be dedicated to understand the implications of this new reality opening up the question if these customizations need to be reflected in RFCs.

\afblock{Acknowledgment.}
Funded by the Excellence Initiative of the German federal and state governments, as well as by the German Research Foundation (DFG) as part of project B1 within the Collaborative Research Center (CRC) 1053 -- MAKI.

\bibliographystyle{IEEEtran}
\bibliography{literature}

\end{document}